\documentclass[11pt]{article}
\usepackage[utf8]{inputenc}
\usepackage{mathtools}
\usepackage[english]{babel}
\usepackage[T1]{fontenc}
\usepackage[font={small,it}]{caption}
\usepackage[a4paper,top=3cm,bottom=2cm,left=3cm,right=3cm,marginparwidth=1.75cm]{geometry}
\usepackage{adjustbox}
\usepackage{rotating}
\usepackage{euscript}

\usepackage{titlesec}

\titleformat*{\section}{\normalsize \bf}
\titleformat*{\subsection}{\normalsize\bf}
\titleformat*{\subsubsection}{\medium\bf}
\titleformat*{\paragraph}{\medium\bf}
\titleformat*{\subparagraph}{\medium\bf}

\usepackage{amsmath}
\usepackage{graphicx}
\usepackage[colorinlistoftodos]{todonotes}
\usepackage[colorlinks=true, allcolors=blue]{hyperref}
\usepackage{wrapfig}

\title{\large \bf Cumulative structure and path length in networks of knowledge}

\author{\normalsize Persoon, P.G.J.\\
\texttt{\normalsize P.G.J.Persoon@tue.nl}
  }
\date{April 2021}

\begin{document}

\maketitle

\begin{abstract}
An important knowledge dimension of science and technology is the extent to which their development is cumulative, that is, the extent to which later findings build on earlier ones. Cumulative knowledge structures can be studied using a network approach in which nodes represent findings and links represent knowledge flows. Of particular interest to those studies is the notion of network paths and path length. Starting from the Price model of network growth, we derive an exact solution for the path length distribution of all unique paths from a given initial node to each node in the network. We study the relative importance of the average in-degree and cumulative advantage effect and implement a generalization where the in-degree depends on the number of nodes. The cumulative advantage effect is found to fundamentally slow down path length growth. As the collection of all unique paths may contain many redundancies, we additionally consider the subset of the longest paths to each node in the network. As this case is more complicated, we only approximate the longest path length distribution in a simple context. Where the number of all unique paths of a given length grows unbounded, the number of longest paths of a given length converges to a finite limit, which depends exponentially on the given path length. Fundamental network properties and dynamics therefore characteristically shape cumulative structures in those networks, and should therefore be taken into account when studying those structures.  
\end{abstract}

\section{Introduction}
\label{Introduction}
Science and technology advance when scientists and inventors learn from earlier findings and use this knowledge to create new findings. A key element of theories of knowledge development is therefore the cumulative nature of discovery and invention \cite{freeman_economics_1997,basalla_evolution_1989,trajtenberg_university_1997,dean_human_2014}, i.e. the building of new knowledge on earlier knowledge. A better understanding of this phenomenon may provide insight into what knowledge development needs to flourish, and how knowledge structures can be built robustly \cite{albert_error_2000, albert_statistical_2002}. Furthermore, a general understanding of cumulative knowledge structures can provide a framework to study how different fields or disciplines of knowledge vary in this dimension, which may help explain variations found across these fields in other knowledge dimensions. In the specific context of technological knowledge, for example, the 'cumulativeness of knowledge' is conjectured to closely relate to the appropriability of that knowledge, as well as to the difficulty by which knowledge travels geographically \cite{nelson_evolutionary_1982,malerba_schumpeterian_1996,breschi_technological_2000}. Understanding how cumulative structures develop is therefore not only relevant from a theoretical perspective, but of great importance as well to targeted science and technology policies aiming to strengthen the development of particular fields.

Approaches to cumulative knowledge structures that aim for a quantitative description may benefit from a network perspective on knowledge. In this perspective, nodes represent findings (which can be any element of knowledge, but usually a scientific finding or an invention) and links represent knowledge connections (indicating that a finding builds on another finding, i.e. knowledge flow in the system). While this may sound abstract, this perspective can, given some limitations\footnote{For example, not all citations may represent knowledge flow. While acknowledging these limitations, we will not go into that discussion here. For an overview in the context of scientific citations see \cite{catalini_incidence_2015,bar-ilan_post_2017} or patent citations see \cite{alcacer_patent_2006, duguet_how_2005}}, be approached empirically using data about publications and citations \cite{garfield_is_1979,price_networks_1965,trajtenberg_penny_1990}. Many contributions studying knowledge networks in this fashion use - or are variations on - a model introduced by Price \cite{price_general_1976}. In this model, nodes are more likely to connect to nodes that already have a large number of knowledge connections, referred to by Price as the 'cumulative advantage effect'\footnote{The term 'cumulative' in this expression, coined by Price, simply means 'added up', and differs from earlier used meaning in 'cumulative knowledge structures', where it suggest the characteristic aspect of knowledge building on knowledge} also known as, in the context of un-directed links, 'preferential attachment' \cite{barabasi_emergence_1999}. In many applications of the Price model, the focus is on degree distributions, which describe how outgoing or incoming links are distributed over nodes \cite{barabasi_emergence_1999, wang_quantifying_2013,steinbock_analytical_2019}. While these distributions to an important extent determine network structures, they are mainly revealing for the variation in the relative importance of nodes, and perhaps less useful to study to what extent there is knowledge flow in such networks. Yet these knowledge flows are an essential element of cumulative structures, in which findings build on findings, which build on other findings, etc. It may therefore be more useful to focus instead on the extent to which sequences of findings appear, which are defined naturally by the well-studied notions of network paths and path length \cite{newman_networks_2010,watts_collective_1998,katzav_analytical_2015}. Yet, where most studies of network paths focus on distance metrics based on considering the \textit{shortest} paths in the network \cite{dereich_typical_2012,dereich_distances_2017,dommers_diameters_2010,caravenna_diameter_2019}, that choice is not at all obvious for knowledge networks. The shortest paths could be misleading in the context of cumulative structures, where one might want to take into account all necessary intermediate steps of development \cite{evans_longest_2020, martinelli_measuring_2014, hu_definition_2011}, which may not be included in the shortest paths.  

\begin{wrapfigure}{l}{0.35\textwidth} 
\centering
\includegraphics[width=0.35\textwidth, height=0.25\textwidth]{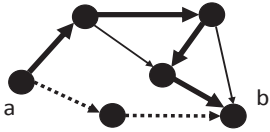}
\caption{\textbf{Types of paths} Between node $a$ and $b$ we can distinguish between the shortest path (dashed links), the longest path (fat links) and all unique paths (the paths formed by dashed, fat or thin links or any combination thereof).}
\label{fig2}
\end{wrapfigure}

As an alternative, one might therefore consider metrics based on the \textit{longest} paths instead (see Figure \ref{fig2}), the length of which necessarily represents the maximum number of intermediate developmental steps. Yet, if we limit the analysis to the longest (or shortest) path between two findings, we ignore that there may be more paths between these findings, which may describe equally relevant sequences of developmental steps. Indeed a key element of invention and discovery is exactly the combination (or sometimes 'recombination') of different ideas \cite{arthur_nature_2009, strumsky_identifying_2015,kaplan_double-edged_2015}, which may be drawn from different sequences of development. To account for these, we may as another alternative consider metrics based on \textit{all unique} paths (for an illustration see Figure \ref{fig2}), for example, the average length of these paths. A downside of considering all paths is that, especially when the average degree is large, there may be many paths between two findings, and not all of these may represent distinct knowledge flows leading to distinct recombined ideas. For example, when two paths leading to a finding largely overlap, the content conveyed in the knowledge flow they represent may largely be the same, and considering them separately is largely a redundant effort. As both alternatives therefore have advantages as well as disadvantages, it may be useful to consider both of them to study cumulative structures.    

It is however not immediately clear how, in the context of knowledge networks, the metrics based on shortest paths can be generalized for the longest paths or all unique paths. Starting from the Price model, Evans et al. make an important contribution, deriving a lower bound for the length of the longest path in a network \cite{evans_longest_2020}. While this is insightful about the longest stretch of knowledge flow in a network, as we argued earlier, there are usually many more paths in a network, some of them representing equally interesting sequences of findings. The longest path with length $l$ might be exceptional, begging the question of how many paths there are of length $l-1$, $l-2$ etc, i.e. how the number of paths is distributed over various lengths. 

A detailed understanding of the path length distributions in knowledge networks allows us to form well-founded expectations of the typical stretch of knowledge flows in cumulative structures and is therefore key to interpret variation in these structures across different scientific disciplines or technologies. In this contribution, we therefore explore the typical path length distributions we might encounter in knowledge networks, and how we can use these to calculate metrics such as the expected path length. We will in a first way of counting network paths consider the distribution of all unique paths coming from a given initial node (see Section \ref{sec1}). Following the Price model, we thereby investigate the role of the cumulative advantage effect. Motivated by recent results which indicate that the average degree, which is usually kept constant, may in fact steadily increase with the number of nodes in a knowledge network \cite{persoon_how_2020}, we consider a generalization of the model allowing for this increase. In a second way of counting paths (see Section \ref{sec2}), we focus on a subset of all unique paths, by selecting only the \textit{longest paths} from the initial node to each node in the network. As deriving an exact solution for this distribution is challenging, we will approximate it instead, thereby ignoring the cumulative advantage effect. Though simplified, this allows us to derive the main characteristics of the distribution, approximate the expected longest path length and compare it to the case of all unique paths. 

\section{All unique paths in the Price model}\label{sec1}

For each discrete step in time $n$, the Price model generates a directed acyclic graph $G(n)$ consisting of $N$ nodes and $M$ links \cite{price_general_1976}. Starting from some initial acyclic graph $G(1)$, at each step in time, a new node is added to the network, which is connected with incoming links to an average of $\langle m \rangle$ existing nodes in the network. The number of incoming links of a node $l$ in $G(n)$, i.e. its in-degree, therefore does not change as $n$ increases, yet the number of outgoing links of $l$, i.e. its out-degree, is however expected to gradually increase with $n$. In the context of knowledge networks, the incoming links of a node $l$ represent the set of knowledge connections appearing at once when $l$ is introduced (i.e. published, patented), hence $l$ can be interpreted to 'build on' the set of nodes to which it is connected by the incoming links. Reversely, the set of nodes to which $l$ is connected by its outgoing links can be interpreted to build on $l$. Note this implies that the links, (and thus the paths), are in the direction of knowledge flow, which is a convention in line with Evans, yet opposite to a number of others \cite{newman_networks_2010, steinbock_analytical_2019,vazquez_statistics_2001}. In most applications of the Price model, it is assumed that the average in-degree $M/N=\langle m \rangle$ is approximately constant as the network grows. 

In this contribution, our initial graph $G(1)$ consists of a single 'initial node' $1$ and we number the subsequent nodes by the order of appearance: $2,3,..., n$, hence at any time, $N=n$. While this choice for an initial graph allows for a simple description of the growth process, it also introduces two subtleties. First, for $n=1$ there are no other nodes to connect to, so insisting that $\langle m \rangle>0$ at that point appears problematic. As an exception, we will allow this node (and only this node) to connect to itself. Second, especially when $n$ is small and $\langle m \rangle$ is large, new nodes may not have enough distinct nodes to connect to. Therefore, we allow multiple linkages to the same node, which should occur more rarely when the network becomes larger. We refer to Evans \cite{evans_longest_2020} for a more elaborate discussion of these subtleties.

In the Price model, the probability $\Pi(n,l)$ for a new node $n+1$ to connect to an existing node $l$ consists of two parts: (i) a part which is non-zero and equal for all nodes, and (ii) a part which is proportional to the out-degree $h(l,n)$ of $l$. Introducing the constant $c\geq 0$ which represents the strength of effect (ii) in proportion to effect (i)\footnote{We note that in the original model of Price, $c=1$ and in the approach by Evans, the parameter $p=cm/(1+cm)$ is instead introduced.}, we can thus write
\begin{equation}
    \Pi(l,n)=\frac{1+ ch(l,n)}{\sum_{t}^{n}1+ ch(t,n)}=\frac{1+ ch(l,n)}{n+c\langle m\rangle n},
\end{equation}
hence note that when $c=0$, the 'cumulative advantage effect' is switched off, and we are left with the neutral case where new nodes link equally likely to any node in $G(n)$. For simplicity we will in this work only consider the paths in $G(n)$ starting from the initial node (in the Section \ref{discussion} we discuss some generalizations of this choice), so when we mention in the following 'a path to node $l$' we mean a unique path from the initial node to node $l$. The number of paths $f_{k}(n)$ are likewise defined as the total number of unique paths of length $k$ in $G(n)$ starting from the initial node. We assume there is a single path of length zero from the initial node to itself, i.e. $f_{0}=1$ for all $n$, though this largely a matter of convention. We will derive an expression for the expected value $\langle f_{k}(n)\rangle$, yet for brevity we drop the $\langle\rangle$ notation, also for $\langle m\rangle$. Let $q_{k}(l)$ be the number of paths to node $l$ with length $k$, hence when a new node connects to $l$, there are $q_{k}(l)$ new paths of length $k+1$. The expected increase in the number of paths of length $k+1$ is therefore
\begin{equation}\label{main}
    \Delta_{n}f_{k+1}(n)=m\sum_{l=1}^{n}q_{k}(l)\Pi(l,n)= m\sum_{l=1}^{n}\frac{q_{k}(l)+cq_{k}(l)h(l,n)}{(1+cm)n}.
\end{equation}
We note that each of the $q_{k}(n)$ paths going through $l$ extend into $q_{k}(n)h(l,n)$ paths of length $k+1$, therefore $\sum_{l}h(l,n)q_{k}(n)=f_{k+1}(n)$ and using that $\sum_{l}q_{k}(l)=f_{k}(n)$, we obtain 
\begin{equation}\label{main2}
    \Delta_{n}f_{k+1}(n)=m\frac{f_{k}(n)+cf_{k+1}(n)}{(1+cm)n}.
\end{equation} 
Additionally, we have the initial condition that $f_{k}(1)=0$ for all $k>0$ as there are no paths of length $k>0$ when $n=1$. Before we discuss the general solution, let us focus briefly on the simple neutral case where we exclude the cumulative advantage effect.

\subsection{Excluding the cumulative advantage effect}\label{secneutral}

Excluding the cumulative advantage effect amounts to setting $c=0$. Equation \ref{main2} then becomes $\Delta_{n}f_{k+1}(n)=mf_{k}(n)/n$. Noting that $f_{0}=1$, this basic relation is directly solved by 
\begin{equation}
    f_{k}(n)=\frac{m^{k}}{\Gamma(n)}S(n,k+1)
\end{equation}
where $\Gamma(n)$ is the gamma function and $S(n,k)$ is the $n,k^{th}$ unsigned Stirling number of the first kind. The latter appear as coefficients in the rising factorial of a real number $x$ to height $n$, defined in mathematics as
\begin{equation}\label{rising}
    x^{\overline{n}}=x(x+1)...(x+n-1)=\sum_{k=0}^{n}S(n,k)x^k.
\end{equation} 
Stirling numbers can be expressed in terms of harmonic numbers and generalized harmonic numbers \cite{adamchik_stirling_1997}, for example allowing us to write for $f_{1}(n)=mH(n-1)$, where $H(n)$ is the $n^{th}$ harmonic number. When $n$ gets large, the leading term of $S(n,k+1)/\Gamma(n)$ is approximately $\log(n)^{k}/\Gamma(k+1)$ \cite{wilf_asymptotic_1993}. For large $n$, the number of paths of length $k$ can for large $n$ therefore be approximated as $m^{k}\log(n)^k/\Gamma(k+1)$, which we can recognize this as a (not normalized) Poisson distribution of the variable $k$. 

Using Equation \ref{rising} we can derive the expected total number of paths $K(n)=\sum_{k}f_{k}(n)$, to equal
\begin{equation}
    K(n)=\frac{\Gamma(m+n)}{\Gamma(n)\Gamma(m+1)}
\end{equation}
This expression increases approximately as $n^m$. To obtain the expected path length $\ell(n)=\sum_{k}kf_{k}(n)/K(n)$ note that we can differentiate $K(n)$ with respect to $m$ and multiply by $m/K(n)$, resulting in 
\begin{equation}
    \ell(n)=m\psi(m+n)-m\psi(m+1),
\end{equation}
where $\psi(m+n)$ is the digamma function, which increases logarithmically in $n$. We conclude therefore that the expected path length of all unique paths increases logarithmically with the number of nodes $n$, along with a coefficient $m$. To be able to compare this relation to later cases we can denote it more generally as 
\begin{equation}\label{defd}
    \ell(n)\approx d_{m} \log(n)+\ell_{1},
\end{equation}
where the coefficient $d_{m}$ is some constant depending on $m$ and $\ell_{1}$ is another constant we are less interested in. For the case where there is no cumulative advantage effect we therefore have $d_{m}=m$.   

\subsection{Including the cumulative advantage effect}

\begin{wrapfigure}{r}{0.6\textwidth} 
\centering
\includegraphics[width=0.6\textwidth, height=0.38\textwidth]{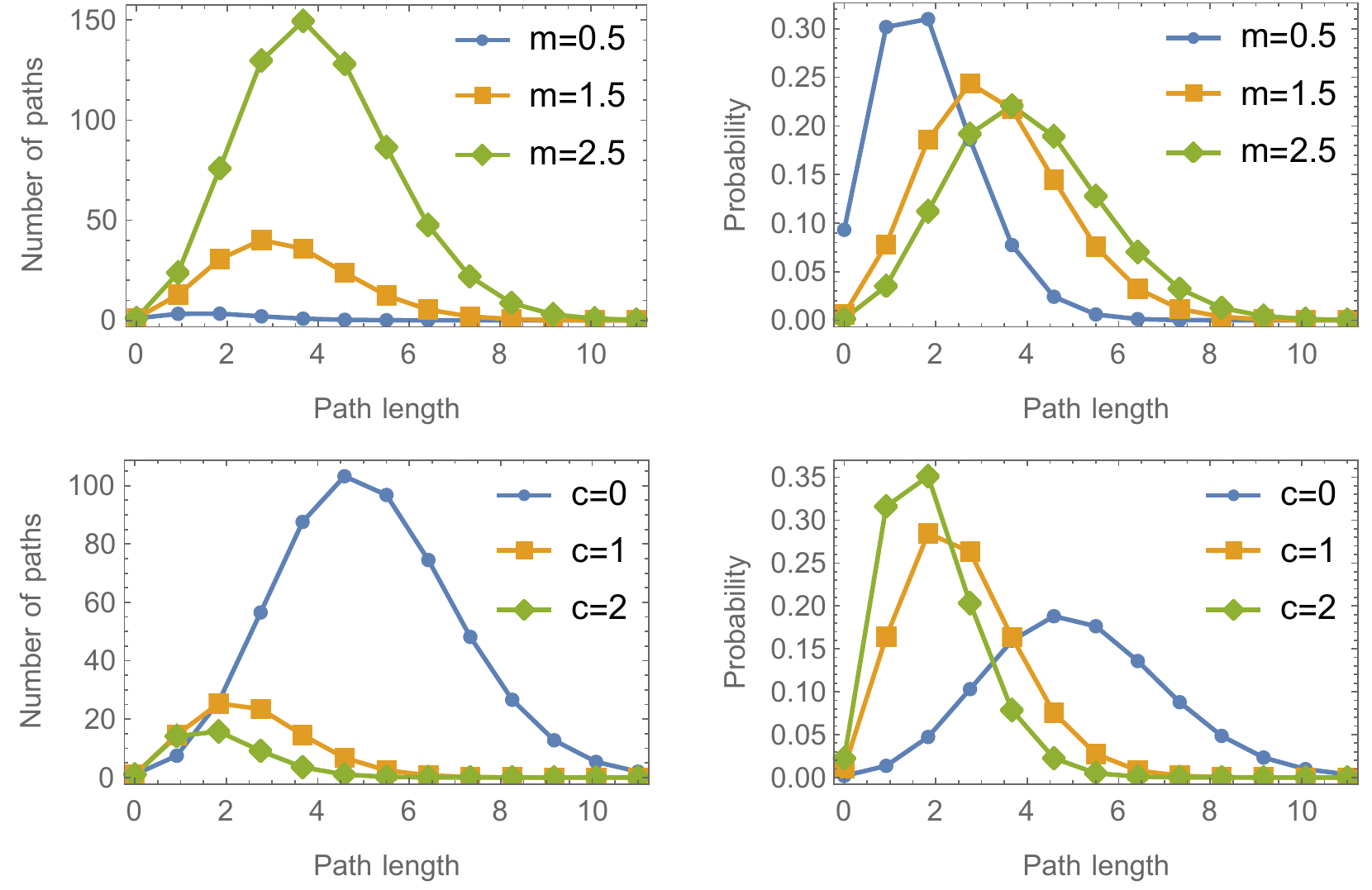}
\caption{\textbf{Distribution of path lengths} In the left two panels we plot the distribution of the number of paths for various values of $m$ and $c$, in the right two panels we plot the same distributions but then normalized for the number of paths. Unless otherwise specified, $n=80$, $m=1.5$ and $c=0.5$. We observe that there are less paths for lower $m$ and greater $c$ and that the distributions are more skewed to lower path length values for lower $m$ and greater $c$.}
\label{fig1}
\end{wrapfigure}

For general values of $c$ the analysis becomes slightly more complicated. Going back to Equation \ref{main2}, let us start by writing down the general solution (we refer to the supplementary material for a detailed derivation):  
\begin{equation}\label{centralsol}
    f_{k}(n)=\frac{1}{\Gamma(n)(-c)^k}\sum_{t=k+1}^{n}(t-k)S_{p}(n,t)(-p)^{t-1}.
\end{equation}
where $p=cm/(1+cm)$ and $S_{y}(n,t)$ is the $n,t^{th}$ non-central unsigned Stirling number of the first kind \cite{koutras_non-central_1982, schmidt_generalized_nodate}, which are defined for any real $y$ by a slight variation of Equation \ref{rising}, namely $x^{\overline{n}}=\sum_{k=0}^{n}S_{y}(n,k)(x-y)^k$ and in particular $S_{0}(n,k)=S(n,k)$. Note that for $c\rightarrow 0$, we have $p\rightarrow 0$, $p/c\rightarrow m$ and the only member in the sum of Equation \ref{centralsol} not going to zero is the first term $(p/c)^{k}S_{p}(n,k+1)\rightarrow m^{k}S(n,k+1)$, thus retrieving the solution for $c=0$. We plot the distribution, for a number of values of $m$ and $c$, in Figure \ref{fig1} (left two panels), including the case $c=0$. We observe the distributions for greater $c$ are more skewed towards lower path length values, it appears therefore that the cumulative advantage effect tempers the path length growth. Specifically considering
\begin{equation}\label{f1}
    f_{1}(n)=\frac{1}{c}\Big(\frac{\Gamma(p+n)}{\Gamma(p+1)\Gamma(n)}-1\Big),
\end{equation}
we see that $f_{1}(n)$ is initially smaller than $mH(n-1)$ (i.e. the value for $f_{1}(n)$ when $c=0$), yet for a given $n$, it will overtake $mH(n-1)$ and subsequently grow much larger. Where in the limit of large $n$, $mH(n-1)$ increases logarithmically, the expression in Equation \ref{f1} increases as $n^{p}$. We can show that the $f_{k}(n)$ for $k>1$ show similar behavior. This leads us to the conclusion that, up to a given length $\mathsf{k}$, there are many more paths when there is a cumulative advantage effect, yet beyond that length $\mathsf{k}$, there are actually fewer paths (compared to the $c=0$ case). In other the words, there tend to be more shorter paths when there is a cumulative advantage effect. Finally, in the supplementary material we show that the leading order of $f_{k}(n)$ for large $n$ can be approximated as 
\begin{equation}
    f_{k}(n)\approx \Big(\frac{p}{c}\Big)^{k}\frac{\Gamma(n+p)}{p\Gamma(1+p)\Gamma(n)\Gamma(k)}\log\Big(\frac{n+p}{1+p}\Big)^{k-1},
\end{equation}
which, up to a factor depending on $n$, we may again recognize as a (not normalized) Poisson distribution of the variable $k$.    

Again summing $f_{k}(n)$ over all $k$, we obtain for the total number of paths 
\begin{equation}\label{centraltot}
    K(n)=\frac{\Gamma(m_{c}+n)}{(1+c)\Gamma(n)\Gamma(m_{c}+1)}+\frac{c}{1+c}, 
\end{equation}
where $m_{c}=m(1+c)/(1+cm)$. For large $n$ we can conclude this expression grows approximately as $n^{m_{c}}$. Note that $m_{c}<m$ for $m>1$, hence the power of $n$ by which the number of paths increase is here smaller than the one derived in the $c=0$ case. In line with the observations with Figure \ref{fig1}, the cumulative advantage effect thus slows down the growth of the number of paths for $m>1$. However, when $0<m<1$, $m_{c}$ is actually larger than $m$, hence, in that case, the cumulative advantage effect somewhat accelerates the growth of the number of paths. This effect, apart from the fact that $m<1$ may be rather uncommon in knowledge networks, is however limited: rewriting $m_{c}$ as $1-\frac{1-m}{1+cm}$, we see that, given $0<m<1$, it will still always be smaller than $1$ for any $c$. Therefore, we conclude that $m$ alone determines whether the number of paths increases faster than linear or not. We can divide $f_{k}(n)$ by $K(n)$ to obtain the normalized path length distributions, which we depict for a number of values in Figure \ref{fig1} (right two panels). In line with the observations for the not-normalized distribution, these plots indicate that the shorter paths are more probable for lower $m$ and greater $c$. 

To obtain the expected path length $\ell(n)$, we show in the supplementary material how $K(n)$ can with a minor adaptation be approached as a generating function, which allows us to straightforwardly calculate $\ell(n)=\sum_{k}kf_{k}(n)/K(n)$, resulting in 
\begin{equation}\label{centralexp}
    \ell(n)=\frac{1+c+m_{c}\psi(m_{c}+n)-m_{c}\psi(m_{c}+1)}{cr(n)+1+c}-\frac{1}{1+c}
\end{equation}
where $r(n)=(K(n)-c/(c+1))^{-1}$. In the limit of large $n$, $r(n)\rightarrow 0$. We can then approximate
\begin{equation}\label{centralexpsimple}
    \ell(n)\approx \frac{m_{c}\psi(m_{c}+n)-m_{c}\psi(m_{c}+1)+c}{1+c}.
\end{equation}
This again shows that the expected path length increases logarithmically in $n$. The only difference with the $c=0$ case is that the coefficient of $\psi(m_{c}+n)$, i.e. $d_{m}$, is here $m_{c}/(1+c)$ instead of $m$. Noting that $m_{c}/(1+c)=m/(1+cm)<m$ for any $m>0$, we conclude that, compared to the $c=0$ case, the cumulative advantage effect slows down the development of the expected path length by a factor proportional to $c$. Furthermore, the cumulative advantage effect puts an upper limit on $d_{m}$ of value $1/c$ (which is reached only for very large in-degree). This upper limit is therefore lower when the cumulative advantage effect is greater. Note that the upper limit on $d_{m}$ disappears only when $c=0$. 

\subsection{Generalization for increasing average in-degree}

Finally we discuss an extension of the model where we allow the average in-degree $m$ to depend on $n$, i.e. considering a number of expressions for $m(n)$. Equation \ref{main} then becomes 
\begin{equation}\label{main3}
    \Delta_{n}f_{k+1}(n)= m(n)\frac{f_{k}(n)+cf_{k+1}(n)}{n+c\sum_{l}^{n}m(l)}.
\end{equation}
When we take $m(n)$ to be any linear combination of integer or non-integer powers of $n$, which is finite and positive for all $n$ and in which the largest power of $n$ has an exponent $\alpha>0$, then in the limit of large $n$, Equation \ref{main3} reduces to
\begin{equation}\label{main4}
    \Delta_{n}f_{k+1}(n)\approx (1+\alpha)\frac{f_{k}(n)+cf_{k+1}(n)}{cn}.
\end{equation}
This equation is similar to Equation \ref{main2} if we make the substitution $(1+\alpha)/c=m/(1+mc)$. For large $n$, we therefore have the same dynamics as in earlier model with $m=-\frac{\alpha+1}{c\alpha}$ (where $c\neq 0$). This substitution may at first seem odd, as when $m$ was interpreted as the average in-degree, it was restricted to positive values. This assumption was used mainly to interpret the results however, and we see that as long as $m_{c}>0$, the derivation leads to the same equations for negative $m$. In fact we obtain perfectly acceptable results when, using $m=-\frac{\alpha+1}{c\alpha}$, we note that the parameter $p$ (appearing in Equation \ref{centralsol}) becomes $\alpha+1$ and $m_c$, (appearing in Equation \ref{centraltot}) becomes $(\alpha+1)(1+c)/c$. Recalling that $m_{c}$ is the power of $n$ by which the total number of paths increase, we thus conclude that the smaller the cumulative advantage effect, the stronger the number of paths increase, but at least by a power $\alpha+1$. For the expected path length we similarly conclude that the coefficient $d_{m}=m_{c}/(1+c)$ appearing in Equation \ref{centralexpsimple} becomes $(\alpha+1)/c$. We therefore conclude the expected path length still increases logarithmically in $n$, yet with a coefficient $d_{m}$ which is (a) proportional to the largest power of $n$ appearing in $m(n)$ and (b) inversely proportional to the strength of the cumulative advantage effect. 

In the above generalization the assumption that $c\neq 0$ is rather crucial. As is shown in detail in \cite{persoon_how_2020}, the situation becomes rather different with $c=0$ and $m(n)\propto n$. The number of paths then increases exponentially in $n$ and the expected path length increases linearly in $n$.\footnote{The approach in \cite{persoon_how_2020} is slightly different: in that contribution we count each path to an increasing number of initial nodes. Yet it can be demonstrated (see supplementary material) that this amounts to a simple change of initial conditions, the effect of which on the number of paths and expected path length is negligible for large $n$.} It can be demonstrated that when $m(n)$ grows faster than linear in $n$ for $c=0$, the number of paths increases even faster than exponentially, and likewise the expected path length increases even faster than linear in $n$. This suggests therefore that the cumulative advantage effect plays a crucial role in keeping the number of paths a power of $n$ and the expected path length a logarithmic relation in $n$, thus fundamentally slowing down the path length dynamics for the case that $m(n)$ increases with $n$. Only when $m(n)$ increases \textit{even} faster in $n$, namely exponentially, the sum appearing in the denominator of Equation \ref{main3} will be proportional to $m(n)$, thus leading for large $n$ to the relation $\Delta_{n} f_{k}(n)\propto f_{k}(n)+cf_{k+1}(n)$, which can be demonstrated to result in expected path length growth linear in $n$. We conclude that, in order to break through the 'logarithmic barrier' imposed by the cumulative advantage effect, the in-degrees need to grow at least exponentially with the number of nodes.

\section{Sub-selecting the longest paths}
\label{sec2}

In Section \ref{sec1} we derived that, when the average in-degree is larger than 1, the number of paths in the network increases rather fast. In the context of knowledge networks, not all of these paths may represent relevant knowledge flows, and there will be many redundancies when each unique path is considered separately. It may therefore make sense to focus instead for each node $l$ on the \textit{longest path} from the initial node to $l$. We will call these paths in the following 'longest paths', yet they should not be confused with the single, unique longest path in the whole network, which is the subject of work by Evans  \cite{evans_longest_2020}. 

Note that the longest path from the initial node to a node $l$ may not be unique. In the following, we will however assume we just choose one longest path from the initial node to each node in $G(n)$ and we are interested in deriving an expression for the number $\mathsf{f}_{k}(n)$ of such longest paths of length $k$. As before we have $\mathsf{f}_{0}=1$ for all $n$. For simplicity we will focus in this derivation on the $c=0$ case and keep $m$ constant, we leave those generalizations for later work. 

We start by noticing that, when a new node $n+1$ connects to a node $l$ in $G(n)$, and there is a longest path from the initial node to node $l$ of length $k$, we necessarily obtain a longest path of length $k+1$ from the initial node to node $n+1$. As there are exactly $\mathsf{f}_{k}(n)$ nodes to which the initial node has a longest path of length $k$, the probability of obtaining a longest path of length $k+1$ to node $n+1$ using one of the links in the in-degree of $n+1$ is $\mathsf{f}_{k}(n)/n$. The probability to create a path with length $k+1$ or less using one the links in the in-degree of $n+1$ is thus $\sum_{t=0}^{k}\mathsf{f}_{t}(n)/n$. Hence collectively considering all links in the in-degree of $n+1$, the probability to obtain a longest path of length $k+1$ is $(\sum_{t=0}^{k}\mathsf{f}_{t}(n)/n)^{m}-(\sum_{t=0}^{k-1}\mathsf{f}_{t}(n)/n)^{m}$, and the expected increase $\Delta_{n}\mathsf{f}_{k+1}(n)$ is
\begin{equation}\label{mainlongest}
  \Delta_{n}\mathsf{f}_{k+1}(n)=\Big(\sum_{t=0}^{k}\frac{\mathsf{f}_{t}(n)}{n}\Big)^{m}-\Big(\sum_{t=0}^{k-1}\frac{\mathsf{f}_{t}(n)}{n}\Big)^{m}  \end{equation}
Introducing $H_{k}(n)=\sum_{t=0}^{k}\mathsf{f}_{t}(n)$ i.e. the number of longest paths with length shorter than $k+1$, and summing both the left and the right of Equation \ref{mainlongest} over $k$, starting from $k=0$, we obtain 
\begin{equation}\label{mainlongest2}
  \Delta_{n}H_{k+1}(n)=n^{-m}H_{k}(n)^{m}.
\end{equation}
It is not straightforward to obtain an exact solution to this equation, we can however identify a number of characteristic properties and use these to derive a greater estimate of $\mathsf{f}_{k}(n)$. First, we rewrite Equation \ref{mainlongest2} as
\begin{equation}\label{mainlongest3}
    H_{k+1}(n)=1+\sum_{s=1}^{n-1}\frac{H_{k}(s)^{m}}{s^{m}}
\end{equation}
From this form, knowing that $H_{0}(n)=1$ for all $n$, it is clear that for $n\rightarrow \infty$ and $m>1$, we have $H_{1}(n)\rightarrow \zeta(m)+1$, where $\zeta(m)$ is the Riemann Zeta function. The number of longest paths of at most length $1$ hence does not grow unbounded, but instead converges to some finite value. In turn we can use the fact that $H_{1}(n)$ converges to show that $H_{2}(n)$ converges, etc., concluding that each $H_{k}(n)$ ultimately converges to some limit for $n\rightarrow \infty$, which we will denote by $H_{k}^{\infty}$. Likewise, the $\mathsf{f}_{k}(n)$ will converge to $\mathsf{f}_{k}^{\infty}=H_{k}^{\infty}-H_{k-1}^{\infty}$. A main question about the distribution is therefore: how does $H_{k}^{\infty}$ depend on $k$? In the following we analyze this relation in more detail by simplifying the dependence of $H_{k}(n)$ on $n$. More precisely, in the next section we discuss a zeroth order approximation in $n$ of $H_{k}(n)$ and in the section that follows a first order approximation in $n$.  

\subsection{Zeroth order approximation}\label{zeroth}

We will investigate the dependence of $H_{k}^{\infty}$ on $k$ by maximally simplifying the dependence of $H_{k}(n)$ on $n$. We start by noting that, since $H_{k}(n)$ counts the number of longest paths of length $k$ or less, we have that $H_{k}(n)=n$ for $n<k$. For $n<k$ therefore, $H_{k}(n)$ increases linear with $1$ path per added node. At the same time, we see from Equation \ref{mainlongest2} that, for $n>k$, $\Delta_{n} H_{k}(n)$ monotonously decreases to zero, hence $H_{k}(n)$ slowly but gradually comes closer to $H_{k}^{\infty}$. We can roughly approximate this development by assuming $H_{k}(n)$ continues to grow linear in $n$ until it reaches $H_{k}^{\infty}$ at $n=H_{k}^{\infty}$, after which it no longer increases and takes the constant value $H_{k}^{\infty}$. As there is no real dependence on $n$ in this approximation, we will refer to this as a zeroth order approximation in $n$. As $H_{k}(n)$ in fact already starts to be slightly smaller than $n$ for $n>k$, we note that our approximation is generally equal or greater than the actual value, hence resulting in an overestimation of $H_{k}^{\infty}$. Equation \ref{mainlongest3} in this approximation becomes 
\begin{align}\label{approx1}
    H_{k+1}^{\infty}& \approx 1+ \sum_{n=1}^{H_{k}^{\infty}}1+\sum_{n=H_{k}^{\infty}}^{\infty}\frac{H_{k}^{\infty m}}{n^{m}}\\
    &\approx 1+H_{k}^{\infty}+H_{k}^{\infty m}\Big(\frac{H_{k}^{\infty(1-m)}}{m-1}+\mathcal{O}(H_{k}^{\infty-m})\Big).
\end{align}
From this last expression we see that $H_{k+1}^{\infty}\approx 1+H_{k}^{\infty}\frac{m}{m-1}$, hence $H_{k}^{\infty}\propto (\frac{m}{m-1})^{k}$. We therefore conclude that the upper bounds of the number of longest paths of length $k$ depend exponentially on $k$, and a first approximation for the base of the exponent is $\beta_{0}=\frac{m}{m-1}$. This base approaches $1$ for larger values of $m$ with a rate $1/(m-1)$. The upper bounds of $\mathsf{f}_{k}(n)$ therefore increase more slowly in $k$ when the average in-degrees are larger. This makes sense, as with larger in-degrees, the creation of longer paths is more likely, hence resulting in relatively less longest paths with short length. As will turn out later however, this approximation to the exponential base could use some improvement. We will lay out the main steps to arrive at this improvement, for the details we refer to the supplementary material. 

\subsection{First order approximation}

In Equation \ref{approx1} we approximate $H_{k}(n)$ by a linear and a constant part. This translates to an $\mathsf{f}_{k}(n)$ which is zero for $n<H_{k}^{\infty}$, and then abruptly $\mathsf{f}_{k}(n)=\mathsf{f}_{k}^{\infty}$ for $n\geq H_{k}^{\infty}$, hence there is no real dynamic dependence on $n$ in that approximation. As an improvement, we could therefore include the first order of $n$ in our approximation for $H_{k}(n)$. As $H_{k}(n)=n$ for $n<k$ let us suppose, as before, that $H_{k}(n)=n$ up to some $n_{k}$ which we specify later. This allows us to split the sum in Equation \ref{mainlongest3} in a part for $1\leq n<n_{k}$ and a part for greater values of $n\geq n_{k}$, and we suppose that $n_{k}$ is sufficiently large such that the latter sum can be well approximated by an integral, leaving us with, for an $n>n_{k}$ 
\begin{equation}
    H_{k+1}(n)=n_{k}+\int_{n_{k}}^{n}\frac{H_{k}(n)^m}{n^m}dn.
\end{equation}
This relation is satisfied to first order in $n$ for
\begin{equation}\label{basisform}  
H_{k}(n) = 
     \begin{cases}
       \text{$n$} &\quad\text{if $n< n_{k}$ }\\
       \text{$a_{k}-\frac{a_{k-1}^{m}}{(m-1)n^{m-1}}$} &\quad\text{if $n\geq n_{k}$} \\
       \end{cases}
\end{equation}
where it counts for the parameters $a_{k}$ that
\begin{equation}\label{transformation}
    a_{k+1}=n_{k}+\frac{a_{k}^{m+1}}{(m+1)a_{k-1}^{m}}-\frac{1}{(m+1)a_{k-1}^{m}}\Big(a_{k}-\frac{a_{k-1}^{m}}{(m-1)n_{k}^{m-1}}\Big)^{m+1}.
\end{equation}
Note therefore that for $n\rightarrow \infty$, we have $H_{k}(n)\rightarrow a_{k}$ and therefore $a_{k}=H_{k}^{\infty}$. 
Next we specify $n_{k}$. From Equation \ref{mainlongest2} we know that $\Delta_{n}H_{k+1}(n)=1$ for $n<k$ and for greater values of $n$ it (slowly) decreases. While we would like to therefore choose $n_{k}$ as close to $k$ as possible, it should also satisfy $\Delta_{n}H_{k+1}(n)\leq 1$. If we take the solution in Equation \ref{basisform} for $k+1$, differentiate with respect to $n$ and substitute $n_{k}$, we obtain an expression for the slope of $H_{k+1}(n)$ at $n=n_{k}$, which is $a_{k}^m/n_{k}^{m}$. The least value for $n_{k}$ we can thus choose while keeping the slope at $n_{k}$ smaller or equal than $1$ is $n_{k}=a_{k}$, which we shall henceforth implement. Note that this value implies that slope of $H_{k+1}(n)$ equals 1, thus ensuring a smooth transition between the part $n<n_{k}$ and the part $n\geq n_{k}$. Similar to the zeroth order approximation, we can show that this first order approximation to $H_{k}(n)$ is generally equal or above its actual value, and will therefore result in an overestimation of $H_{k}^{\infty}$.

While obtaining an exact solution for the relation in Equation \ref{transformation} after substituting $n_{k}=a_{k}$ remains challenging, we note all the terms on the right-hand side of the equation are of net order $1$ in $a_{k}$ and/or $a_{k-1}$, indicating that, at least for large $k$, $a_{k}$ and thus $H_{k}^{\infty}$ increases exponentially in $k$. Let us suppose for large $k$ we can write $a_{k-1}\beta_{1}=a_{k}$, Equation \ref{transformation} then reduces to 
\begin{equation}\label{betaapprox}
 \beta_{1}=1+\frac{\beta_{1}^{m}}{m+1}-\frac{\beta_{1}^{m}}{m+1}\Big(1-\frac{1}{(m-1)\beta_{1}^{m}}\Big)^{m+1}   
\end{equation}
While this equation does not allow us to write $\beta_{1}$ in terms of elementary functions of $m$, expanding the part in brackets to second order in $1/\beta^{m}$ gives 
\begin{equation}
    \beta_{1}\approx 1+\frac{1}{m-1}-\frac{m}{2(m-1)^{2}\beta_{1}^m}+... 
\end{equation}
This shows that $1<\beta_{1}<\beta_{0}$ for all $m$, thus confirming this approximation is an improvement to the zeroth order approximation. Also, the last term on the right-hand side is of net order $1/(m-1)$, which implies that the term of order $1/(m-1)$ in the expansion of $\beta_{1}$ cannot simply be taken to equal $1/(m-1)$ (as it is for $\beta_{0}$). This indicates that this first order approximation to $H_{k}(n)$ is not just an improvement in orders greater than $1/(m-1)$. In the supplementary material it is demonstrated how may use Equation \ref{betaapprox} to derive the greater estimate 
\begin{equation}\label{betaapprox2}
   \beta_{1}=1+\frac{e-e^{1-\frac{1}{e}}}{m-1}+\frac{e^{-1-\frac{1}{e}} \left(1+3 e+3 e^2\right)-3e}{2(m-1)^2}+...
\end{equation}
This therefore shows that, similar to $\beta_{0}$, the exponential base $\beta_{1}$ approaches $1$ for larger values of $m$, yet where $\beta_{0}$ does so by a rate $1/(m-1)$, $\beta_{1}$ does so by a rate which is an approximate factor $e-e^{1-\frac{1}{e}}\approx 0.84$ smaller.

\subsection{Expected path length}

Theoretically, as long as we can find solutions for $H_{k}(n)$ to second, third etc order in $n$ we can continue to derive better approximations $\beta_{2}$,$\beta_{3}$ etc for the exponential base of $H_{k}^{\infty}$. In this contribution we however stop here and instead derive what the exponential dependence of $H_{k}^{\infty}$ implies for the expected path length. Even though we only demonstrated that $H_{k}^{\infty}$ approaches a exponential function for larger values of $k$, let us approximate $a_{k}\propto\beta_{1}^{k}$ for all $k$. Using the condition that for all $n$ we have a single path of length $0$, thus $H_{0}^{\infty}=\mathsf{f}_{0}^{\infty}=1$, we can approximate
\begin{equation}
   H_{k}^{\infty}=\frac{\beta_{1}^{k+1}-1}{\beta_{1}-1} \quad \text{and} \quad \mathsf{f}_{k}^{\infty}=\beta_{1}^{k}.
\end{equation}
Let us define $k_{n}$ as the largest $k$ for which $\mathsf{f}_{k}(n)$ is non-zero. As we require that the sum $\mathsf{f}_{k}(n)$ over all $k$ to equal $n$, (as there is one longest path for each node), this allows us to write $n=\sum_{s=0}^{k_{n}}\mathsf{f}_{s}(n)= H_{k_{n}}(n)$, or 
\begin{equation}
    n=\frac{\beta_{1}^{k_{n}+1}-1}{\beta_{1}-1}-\Big(\frac{\beta_{1}^{k_{n}}-1}{\beta_{1}-1}\Big)^{m}\frac{1}{(m-1)n^{m-1}}.
\end{equation}
When $n$ gets large the second term on the right-hand side goes to zero (and note this term is absent in the zeroth order approximation of Section \ref{zeroth}). For both the zeroth and first order approximation we can therefore deduce that, when $n$ is large $k_{n}\approx \frac{\log(n(\beta_{1}-1)+1)}{\log(\beta_{1})}-1$, allowing us to compactly write for the distribution
\[   
\mathsf{f}_{k}(n) = 
     \begin{cases}
       \text{$\beta_{1}^{k}$} &\quad\text{if $k\leq\frac{\log(n(\beta_{1}-1)+1)}{\log(\beta_{1})}-1$ }\\
       \text{$0$} &\quad\text{if $k>\frac{\log(n(\beta_{1}-1)+1)}{\log(\beta_{1})}-1$}. \\
       \end{cases}
\]
We can use this to calculate the expected longest path length $\ell(n)=\sum_{k}k\mathsf{f}_{k}(n)/n$. For large $n$ this expression can be shown to reduce to, up to constant terms, $\ell(n)\approx k_{n}$. We therefore conclude that the expected longest path length increases logarithmically in $n$, with a coefficient $d_{m}=\log(\beta_{1})^{-1}$ (see also Equation \ref{defd}), which implies that a greater estimate of $\beta_{1}$ results in a lower estimate of $d_{m}$. Before we consider the value of $d_{m}$ for $\beta_{1}$ in more detail, let us first consider it for $\beta_{0}$ instead. We can for $m>2$ approximate $\log(\beta_{0})^{-1}\approx m-\frac{1}{2}$. Recall that we derived the exact value $d_{m}=m$ when we considered all unique paths in Section \ref{secneutral}. While the value for $d_{m}$ based on $\beta_{0}$ is thus of the same proportion, the small shift of $1/2$ in fact makes it somewhat smaller. $\beta_{0}$ Should therefore not be considered an accurate approximation: the longest paths are expected to be at least as long, yet probably longer on average, than the collection of all unique paths. Hence let us finally approximate $d_{m}$ based on $\beta_{1}$. We thereby use the greater estimate for $\beta_{1}$ 
from Equation \ref{betaapprox2}. This gives $\log(\beta_{1})^{-1}\approx 1.2m-0.6$. Note that, up to a minor shift, this is a factor 1.2 greater than the $d_{m}$ found for all unique paths, which makes more sense than results based on $\beta_{0}$. It suggests that, regardless of the number of nodes and average in-degree in the network, the longest paths are larger than the rest of the paths at least by a fixed proportion. We derived a lower estimate for this of $1:1.2$, yet with an improved approximation of the exponent base $\beta_{1}$ we are likely to find a greater value for this proportion.   

\section{Conclusions}
\label{conclusion}

Studying cumulative structure in knowledge networks is key to understanding the advancement of science and technology, and has besides theoretical implications also relevance for science and technology policies. Approaching a body of knowledge as a network of discrete findings connected through knowledge flows, the notion of network paths and path length can be used to study to what extent sequences of findings appear, which form a key element of cumulative knowledge structures. It is in that context key to study (all) intermediate steps of development, hence not to limit the analysis to the shortest paths. In this contribution, we have therefore studied the path length distribution of (i) all unique paths from a given initial node to each node in the network and (ii) the longest paths from the initial node to each node in the network. 

In the part of this work where we considered all unique paths, we derived an exact solution for the path length distribution and expected path length in the particular context of the commonly used Price model. In this model, two main properties play a role: the average in-degree (AID) and the 'Cumulative Advantage Effect' (CAE). We find that, for large networks, the path length distributions can be characterized as Poisson-like, and are more skewed to lower path length values when the AID is smaller and the CAE is stronger. Similarly, we find that the expected path length grows logarithmically with the number of nodes and that the coefficient of this growth is smaller when the AID is smaller and the CAE stronger. In fact, the CAE puts an upper limit to this coefficient, and this upper limit is lower when the CAE is stronger. The upper limit disappears when there is no CAE. These results are more nuanced when the AID is less than $1$ (which, though possible, may be rather uncommon in knowledge networks). In that case, a stronger CAE may slightly accelerate the growth of the number of paths, yet still has a tempering effect on the path length growth. 

These results may be generalized by allowing the AID to increase with the number of nodes in any power relation. As it turns out, the CAE then plays a crucial role in keeping Poisson-like path length distributions and logarithmic expected path length growth. Only when the AID increases very fast, to be precise exponentially with the number of nodes, then we obtain binomial-like path length distributions and linear path length growth. Without the CAE, these types of path length distribution and expected path length growth would already be obtained for an AID that increases linearly with the number of nodes. The CAE therefore categorically tempers path length growth.

In the part of this work where we consider only the longest paths from an initial node to each node in the network, we only approximate the path length distribution and expected path length, as deriving exact solutions is in this case analytically more challenging. For simplicity, we also focus on the neutral case where the CAE is absent. Notwithstanding our analysis indicates key differences with the case where we consider all unique paths. Where for the latter, the number of paths of a given length grows unbounded, the number of longest paths of a given length is bound to an upper limit. Our approximation suggests that these upper limits increase exponentially with associated lengths and that the base of this exponent is a number slightly larger than 1, and approaches 1 for a greater AID. This makes sense as with a greater AID, we obtain longer paths at a rather earlier stage of the network development than for lower AID. While the distributions over the path lengths thus appear to be rather different, the expected path length appears to develop in fact rather similar. First-order estimates indicate that the expected path length of the longest paths increases at least logarithmically with the number of nodes, with a coefficient proportional to the AID, and an additional constant factor of at least $1.2$. This is similar to the case of all unique paths without the CAE, except for the constant factor of $1.2$. This is however a first theoretical approximation of this factor, and more elaborate approximations are likely to correct this to a greater value. 

To conclude, we have shown that fundamental network properties and dynamics characteristically shape elements of knowledge networks that we can associate with cumulative structures, such as the notion of path length. In particular, the (development of the) AID and the strength of the CAE are relevant properties to consider in this context, as they can be meaningfully interpreted to determine variations in cumulative structures across different knowledge networks. 

\section{Discussion}\label{discussion}

Finally, we discuss some deeper implications and shortcomings of our analysis. First, our results have a number of deeper implications in particular for the study of cumulative knowledge structures. While researchers aiming for a quantitative approach benefit from a network approach to knowledge structures, they should be aware of the various choices that network analysis allows to identify knowledge flow, in particular the differences between using the shortest, longest, or all unique paths in the network. Where the average distance based on the length of the shortest paths in a scale-free network (of which the Price network is a special case) is known to increase with the $\log\log$ number of nodes \cite{cohen_scale-free_2003}, we have shown that the average path length based on the length of all unique paths from an initial node to each node in a Price network increases with the $\log$ number of nodes. Furthermore, we have shown that there are fundamentally different properties of the path length distributions of all unique paths and the subset of longest paths, even without including sophisticated dynamic principles such as the cumulative advantage effect. 

Additionally, before a certain path length metric is applied to study the cumulative structure of a particular field of knowledge or discipline, the researcher is advised to investigate a number of characteristics of the network, such as a possible development of in-degree as the network grows as well as the presence of the cumulative advantage effect. Our work indicates that the presence of either (and especially the presence of both) greatly affects cumulative structures in those networks. Our work allows the researcher to then formulate a number of specific expectations, especially for the path length distribution and expected path length of all unique paths. Our contribution thus provides a first step towards a framework in which cumulative structures can generally be studied and in which variations between fields or disciplines can meaningfully be interpreted. 

A second deeper implication of our results is of more theoretical nature. In this contribution, we have shown that the cumulative advantage effect explicitly prohibits path lengths to grow faster than logarithmically as long as the average in-degree does not increase exponentially. In another contribution \cite{persoon_how_2020}, where we include an empirical analysis of technological knowledge using patent and patent citation data, we actually find that the average path length (based on counting all unique paths) increases linearly, even though the in-degrees do not increase exponentially (but linearly instead). This may imply that the cumulative advantage effect plays no role in these networks, yet, interestingly, other contributions have suggested that the cumulative advantage effect does play a role in these networks \cite{erdi_prediction_2013,valverde_topology_2007}. Another explanation may be that this differs per technology, or that there may be other effects at work, which were not included in this analysis. 

One of those excluded effects, which brings us to the first shortcoming of this analysis, is the time dependence of knowledge dynamics. As other contributions have indicated, these effects may play a rather substantial role \cite{garavaglia_dynamics_2017, golosovsky_power-law_2017}. Indeed one of the criticisms of the Price model is that the oldest nodes in the networks effectively gain the greatest out-degree. In real-life situations, the fact that a finding is old need not automatically imply it is more relevant than any new finding. The model discussed in this work would therefore benefit from an extension which takes into account time effects, such as the fading of relevance. While a number of such models can be found in the literature \cite{golosovsky_power-law_2017, wu_generalized_2014, wang_quantifying_2013}, it is however not directly clear how to analytically calculate the path length distributions in these models. 

A second shortcoming is our focus on (only) counting the paths from a single given initial node. While this focus may be perfect for studies interested in the particular impact or role of a single finding, for a general understanding of cumulative structures, depending too much on a particular choice for a single node might appear arbitrary and may even be misleading. A simple way to generalize this would be to allow for the possibility of multiple initial nodes, or for the number of initial nodes to increase as the network grows. We explain in more detail in the supplementary material and in \cite{persoon_how_2020}, how these choices could be implemented by slightly changing the initial conditions for Equations \ref{main2} and \ref{main3}. While these changes introduce an extra parameter, they are found not to lead to fundamentally different results when we consider networks with a substantial number of nodes. 

\section{Acknowledgements}
The author is grateful to Floor Alkemade, Rudi Bekkers and Elena Mas Tur for helpful comments on the script. This work was supported by NWO (Dutch Research Council) grant nr. 452-13-010.
 
  \bibliographystyle{acm}
  \bibliography{Longest_paths.bib}

\appendix

\end{document}